\documentclass[useAMS]{mn2e}

\def \nhu {cm$^{-2}$}
\def \cs {c \ s^{-1}}

\def \cha {Chandra}
\usepackage{graphicx}

\title[RR Pic (1925): A $Chandra$ X-Ray View]{RR Pic (1925): A $Chandra$
X-ray
View}
\author[Pek\"on Y. and Balman, \c{S}.]{Y.
Pek\"on$^{}$\thanks{E-mail:yakup@astroa.physics.metu.edu.tr} and \c{S}.
Balman$^{}$\thanks{E-mail:solen@astroa.physics.metu.edu.tr} \\ \\
$^{}$Department of Physics, Middle East Technical University,
In\"on\"u Bulvar{\i}, Ankara, 06531, Turkey}

\begin{document}

\maketitle

\label{firstpage}

\begin{abstract}

We present the $\cha$ ACIS-S3 data of the old classical nova RR
Pic (1925). The source has a count rate of
0.067$\pm$0.002 $\cs$ in the 0.3-5.0 keV energy range. We detect the
orbital period of the underlying binary system in the X-ray
wavelengths.  We  also find that the neutral Hydrogen column density
differs for orbital
minimum and orbital maximum spectra with values
(0.25$^{+0.23}_{-0.18}$)$\times$10$^{22}$ \nhu and 
(0.64$^{+0.13}_{-0.14}$)$\times$10$^{22}$ \nhu at 3$\sigma$\ confidence 
level. The X-ray spectrum of RR Pic can be represented by a
composite model of bremsstrahlung with a photoelectric absorption,
two absorption lines centered around 1.1-1.4 keV and 5 Gaussian
lines centered at emission lines around 0.3-1.1 keV
corresponding to various transitions of
S, N, O, C, Ne and Fe . The
bremsstrahlung temperature derived from the fits range from 0.99 to
1.60 keV and the unabsorbed X-ray flux is found to be
(2.5$^{+0.4}_{-1.2}$)$\times$10$^{-13}$ erg cm$^{-2}$ s$^{-1}$ in the 
0.3-5.0 keV range with a luminosity of (1.1$\pm$0.2)$\times$10$^{31}$ erg 
s$^{-1}$ 
at 600 pc. We also detect excess emission in the spectrum possibly 
originating from the reverse shock in the ejecta.
A fit with a cooling flow plasma emission model show enhanced
abundances of He, C, N, O and Ne in the X-ray emitting region
indicating existence of diffusive mixing.

\end{abstract}


\section{Introduction}

Cataclysmic Variables (CVs) are interacting binary systems hosting a
main-sequence secondary (sometimes a slightly evolved star) and a
primary component, a white dwarf (WD) (Warner 1995). The accretion
process occurs mainly through an accretion disc in cases where the
WD does not have substantial magnetic field to channel the accretion
flow onto the WD via the field lines. These systems are typically
categorized by their eruptive behavior like Classical Novae,
Recurrent Novae and Dwarf Novae (Z Cam type, U Gem Type, SU UMa type
etc). In nonmagnetic systems X-rays are produced in the shocks at
the inner disc boundary layer (Patterson $\&$ Raymond 1985; Verbunt
et al. 1997; Baskill, Wheatley $\&$ Osborne 2005). These systems
show hard X-ray spectra with 3-10 keV temperatures. In magnetic CV
systems (mCVs) accreting matter is funneled by the magnetic field of
the WD onto the magnetic poles forming a strong shock.  The mCVs are
classified into two categories where synchronized/almost
synchronized systems are called Polars (Cropper 1990; Ramsay $\&$
Cropper 2004) with magnetic fields in a range of 10-530 MG, these
systems are characterized as being discless. The spectra of Polars
may show the hot (above 10 keV temperature) hard X-ray tails if the
accreting flow is dense and the magnetic fields are low ($\sim$10$^
{-(10-11)}$ M$_{\odot}$ yr$^{-1}$) whereas more rarified streams and
stronger magnetic fields favor cyclotron cooling where the hard
X-ray tails are suppressed. Soft X-ray emission (via reprocessing)
is enhanced once the accretion rate is $\ge$10$^{-9}$ M$_{\odot}$ 
yr$^{-1}$
(Schmidt et al. 2005). Highly  asynchronous systems are called
Intermediate Polars (Patterson 1994; Hellier 1996; Norton, Wynn \&
Somerscales 2004 and references therein) with magnetic fields in a
range less than 10 MG where a truncated disc exists. These systems
show hard X-ray bremsstrahlung spectra with 10-30 keV temperatures.

Classical nova RR Pic had an outburst in 1925 as
a slow nova (expansion speed $\sim$ 400 km s$^{-1}$).
The shell shows "equatorial ring and polar cap/blob"
geometry.
There are similarities and important differences
between the spectra in the ring and blob regions (in C and O lines)
with a shell size of
30$^{\prime\prime}$x21$^{\prime\prime}$ and expansion rate
of 850 km s$^{-1}$ for the ring (Gill $\&$ O'Brien 1998).
The distance of the nova is measured
to be 600$\pm$60 pc (Gill $\&$ O'Brien 1998).

The point source RR Pic has an orbital period of P$_{orb}$$\sim$
0.$^d$14502545(7) (Kubiak 1984). A different periodicity of 15 min
is detected by Kubiak (1984) and accounted for the white dwarf
period; hence making the source a candidate for intermediate polars.
However, Haefner and Schoembs (1985) could not find the 15 min
period with high-resolution photometry and concluded that the 15 min
period is a transient event in the disc rather than the period of
the white dwarf. Warner (1986) also confirms the absence of this 15
min period. Additionally he also finds flickering activity
independent of the orbital phase coming from the disc itself rather
than the hot spot and the system. Another period of 0.1577 days is also 
found interpreted as the superhump period of the
system (Schmidtobreick et al. 2006). Furthermore, the source has a
hardness ratio similar to Polars and the source spectrum in general
differs compared with the non-magnetic CVs (van Teeseling, Beuermann
$\&$ Verbunt 1996). Polarization measurements indicate the existence
of two components of emission, one associated with a hot spot in the
disc and the other in the preceding side of the disc opposite of the
hot spot (Haefner $\&$ Metz 1982). Kubiak (1984) stated that the
main optical light source in the system and the optical eclipse is
due to the eclipse of the hot spot by the secondary. However,
Schmidtobreick, Tappert $\&$ Saviane (2003) conclude that the eclipse
is due to occultation of the emission from the preceding side rather
than the hot spot.

\begin{figure*}
\centerline{
\includegraphics[scale=0.35,angle=270]{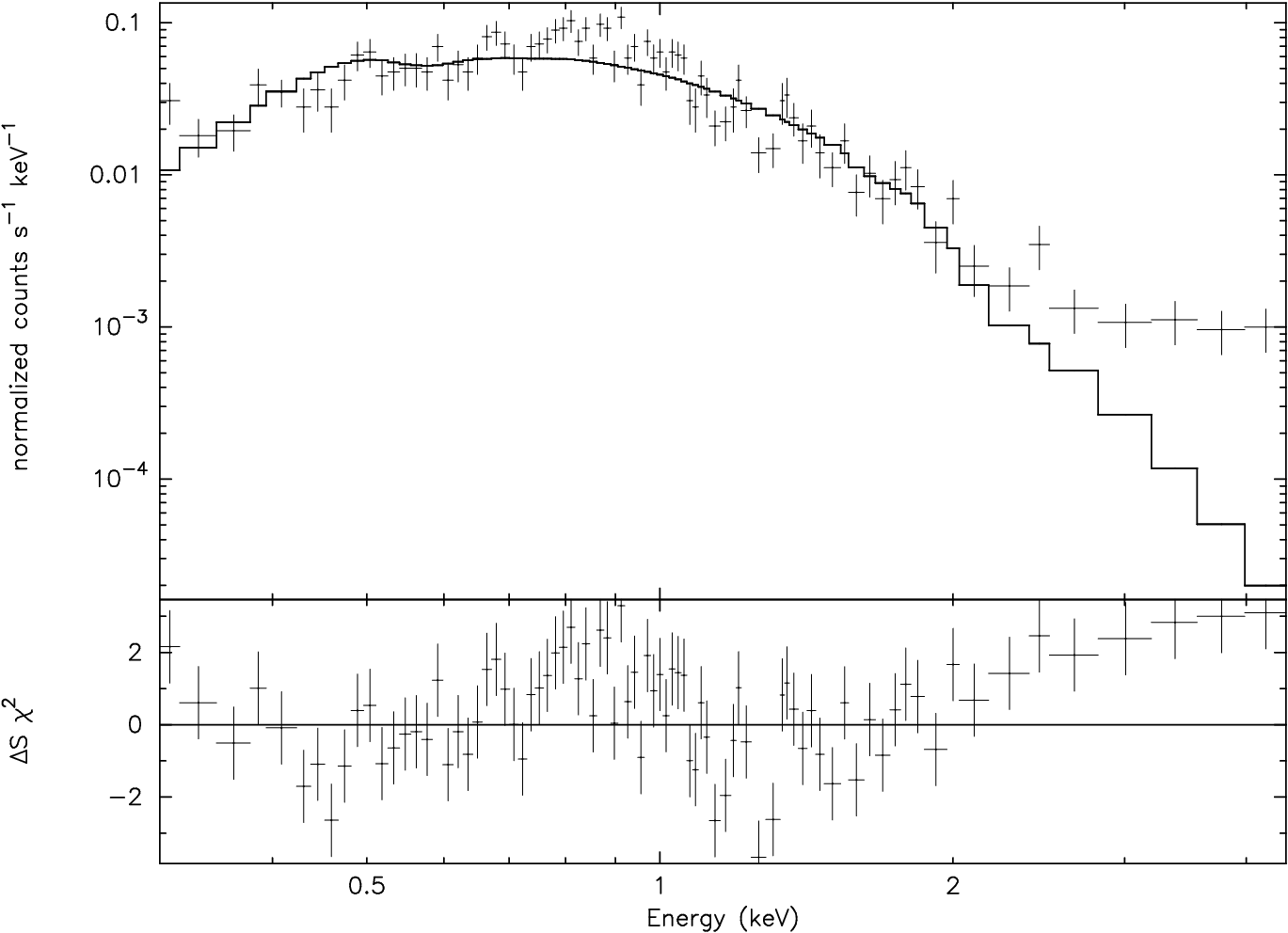}
\includegraphics[scale=0.35,angle=270]{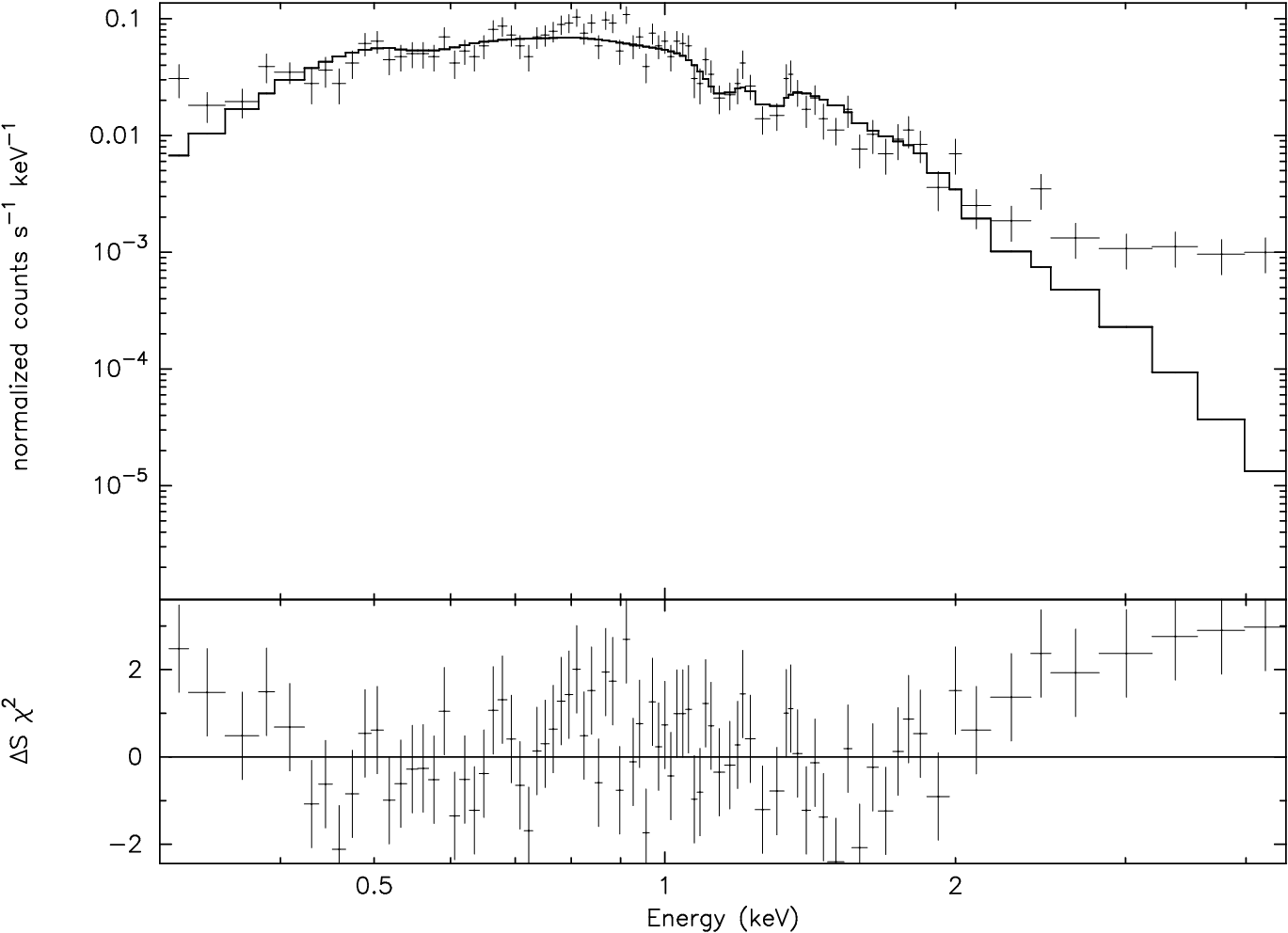}}
\centerline{
\includegraphics[scale=0.35,angle=270]{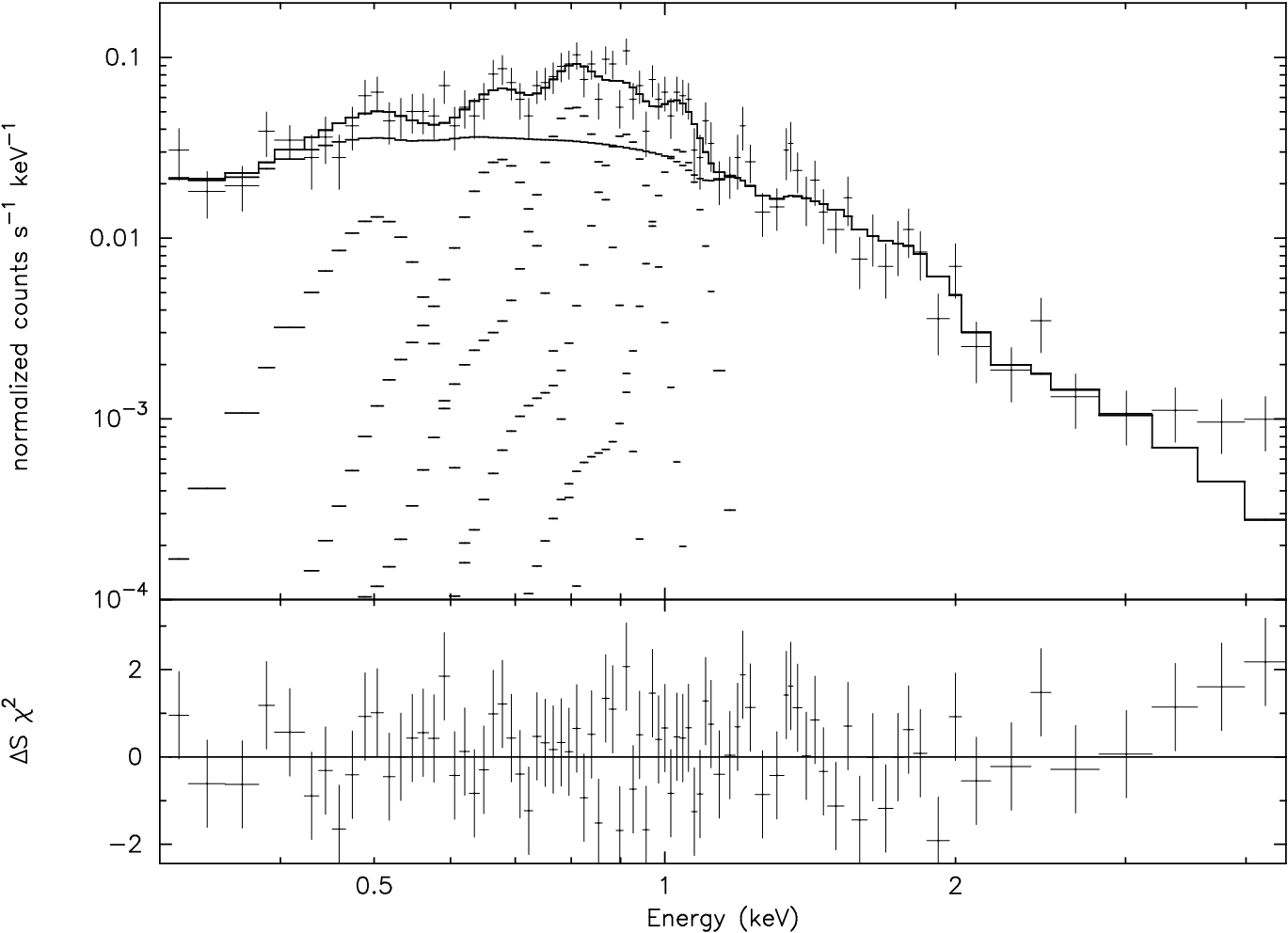}
\includegraphics[scale=0.35,angle=270]{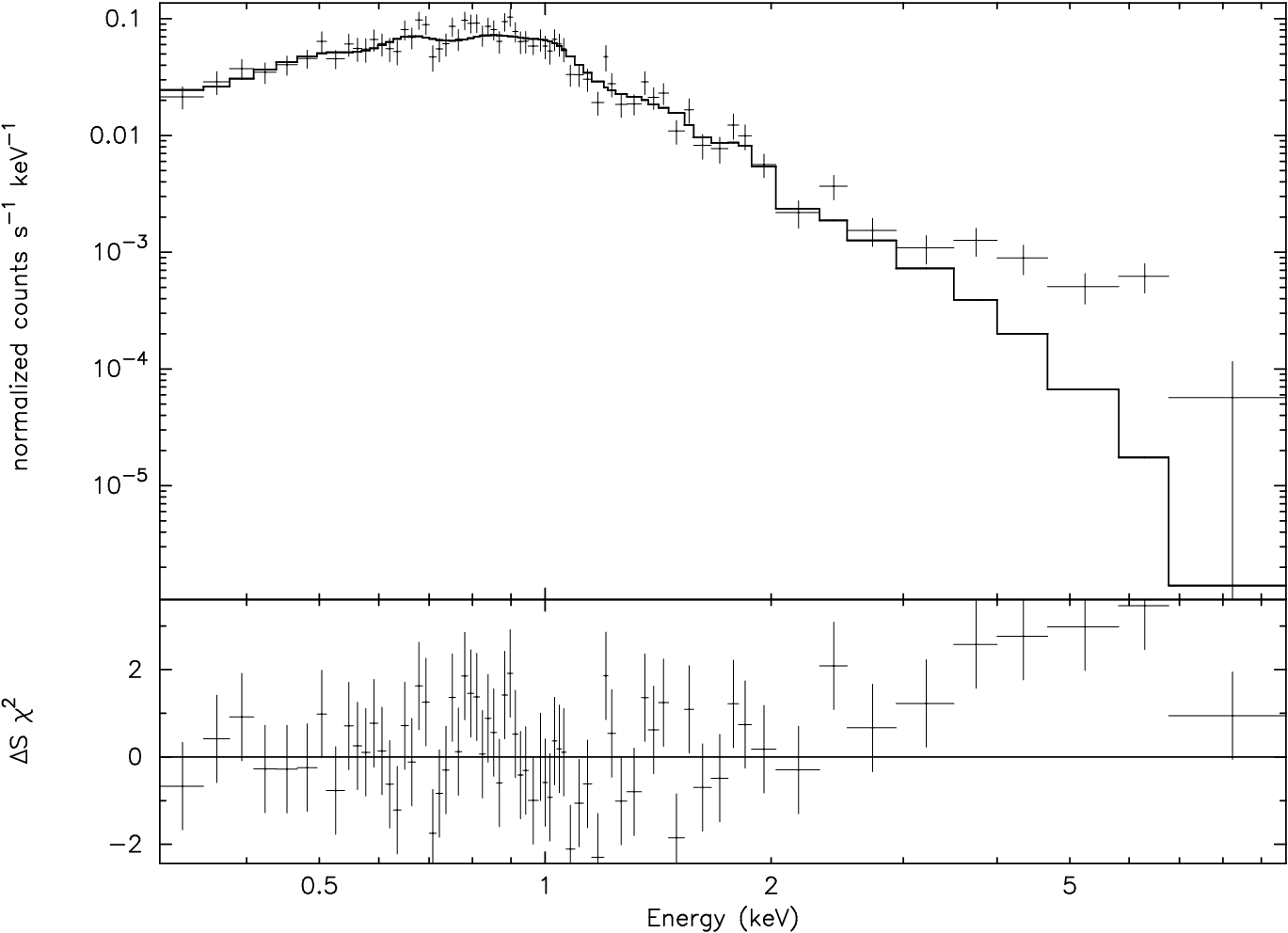}}
\caption{Various spectral fits to the RR Pic data in the 0.3-5 keV range.
The crosses show the data with error bars, solid lines the composite
fitted model and the dashed lines show the individual models.
The panels under the spectra show residuals in standard deviations. The
models fitted to the spectra are as follows: a) Simple Bremsstrahlung, top
left-hand side panel
b) Bremsstrahlung with 2 absorption lines, top right-hand side panel  
c) Bremsstrahlung with 2
absorption and 5 emission lines, bottom left-hand side panel
and  d) VMCFLOW, bottom right-hand side panel. }
\end{figure*}

\begin{table*}
\centering
\begin{minipage}{140mm}
\label{1} \caption{Spectral parameters of the entire spectrum of the
RR Pic in the energy range 0.3-5 keV. $N_{H}$ is the absorbing
column; $kT_{Bremss}$ is the bremsstrahlung temperature; LowT and HighT are low and high temperature values for
VMCFLOW; He, C, N, O, Ne are the abundance ratios with respect to the
solar abundances;  Gabs LineE is the absorption line
centers for the
absorption lines, (the sigma and Tau parameters are frozen at 0.005
and 50 for the first line and 0.01 and 20 for the second line),
Gaussian LineE is the line center for the emission lines (the sigma
values for the lines are frozen at 0.001); $K_{Bremss}$, $K_{VMCFLOW}$ 
and $K_{G}$ are the normalizations for the
bremsstrahlung, VMCFLOW  and Gaussian models respectively. The fluxes
are
given for the entire model in the first row and then for each of the
components in the following rows. All error ranges are given in $\%$90
confidence level ($\Delta \chi^2$=2.71 for a single parameter) }

\begin{center}
\begin{tabular}{@{}lrrrrlrlr@{}}
\hline
\hline

\multicolumn{1}{l}{ }  &
\multicolumn{1}{r}{BREMSS+2ABS+5GAUSS} &
\multicolumn{1}{r}{VMCFLOW} \\

\hline

$N_{H}$ ($\times$ $10^{22}$ atoms/cm$^{2}$) &
0.008$^{+0.014}_{>}$ & 0.008$^{+0.008}_{-0.007}$ \\
$kT_{Bremss}$ (keV) & 1.3$^{+0.3}_{-0.3}$  & N/A1 \\
LowT (keV) & N/A  & 0.14$^{+0.10}_{>}$ \\
HighT (keV) &  N/A & 1.8$^{+0.2}_{-0.2}$ \\
$K_{Bremss}$ & 0.00097$^{+0.00003}_{-0.00002}$ &  N/A \\
$K_{VMCFLOW}$ ($\times$ 10$^{-9}$) & N/A  & 1.8$^{+0.1}_{-0.1}$ \\
He & N/A & 18.4$^{+9.1}_{-6.1}$ \\
C & N/A & 1.7$^{+16.5}_{>}$ \\
N & N/A & 8.7$^{+7.1}_{>}$ \\
O & N/A & 1.9$^{+1.1}_{-0.7}$ \\
Ne & N/A & 1 (frozen)\\
Gabs LineE (keV) & A1:
1.14$^{+0.03}_{-0.02}$ & N/A  \\
& A2: 1.28$^{+0.02}_{-0.02}$ & \\
Gaussian LineE (keV) & G1:
0.53$^{+0.05}_{-0.09}$ & N/A \\
   & G2: 0.66$^{+0.04}_{-0.02}$  &  \\
   & G3: 0.80$^{+0.02}_{-0.02}$  &  \\
   & G4: 0.90$^{+0.02}_{-0.02}$  &  \\
   & G5: 1.02$^{+0.02}_{-0.02}$  &   \\
$K_{G}$ ($\times$ 10$^{-6}$)& G1:
7.3$^{+4.4}_{-5.1}$ &  N/A \\
   & G2: 9.4$^{+4.1}_{-4.1}$  &  \\
   & G3: 14$^{+2}_{-4}$  &  \\
   & G4: 8.4$^{+2.8}_{-2.4}$  &  \\
   & G5: 6.2$^{+1.9}_{-2.0}$  &   \\
Flux ($\times$ $10^{-13}$ ergs/cm$^{2}$/s) &
2.5$^{+0.4}_{-1.2}$ &
2.2$^{+0.3}_{-0.6}$ \\
  &  Bremss: 0.4$^{+0.46}_{-0.28}$  \\
  & G1: 0.07$^{+0.09}_{-0.06}$ \\
  & G2: 0.11$^{+0.08}_{-0.09}$ & & \\
  & G3: 0.19$^{+0.06}_{>}$ & & \\
  & G4: 0.13$^{+0.08}_{-0.07}$ & & \\
  & G5: 0.11$^{+0.10}_{-0.08}$ & & \\
$\chi^2_{\nu}$ & 1.13 (73 d.o.f.) & 1.58 (66 d.o.f) \\

\hline
\end{tabular}
\end{center}
\end{minipage}
\end{table*}
\vspace{0.3cm}

\section{Observation and Data}

RR Pic and its vicinity was observed using the $\cha$ (Weisskopf,
O'dell $\&$ van Speybroeck) Advanced CCD Imaging Spectrometer (ACIS;
Garmire et al. 2003) for a 25 ksec on 2001 October 30 (PI=S.
Balman), pointed at the nominal point on S3 (the back-illuminated
CCD) with no gratings in use yielding a moderate non-dispersive
energy resolution. The data were obtained at the FAINT mode. $\cha$
has two focal-plane cameras and two sets of transmission gratings
that can be inserted in the optical path (HETG and LETG; High and
Low Energy Transmission Gratings). The ACIS is used either to take
high resolution images with moderate spectral resolution or is used
as a read out device for the transmission gratings. ACIS is
comprised of two CCD arrays, a 4-chip array, ACIS-I; and a 6-chip
array, ACIS-S. The ACIS-S3 has a moderate spectral resolution
E/$\Delta$E $\sim$ 10-30 (falls to about 7 below 1 keV) with an
unprecedented angular resolution of 0.49$^{\prime\prime}$ per pixel
(half-power diameter). The standard pipeline processing was done by
CXC. For the spectral and temporal analysis of the data we used the 
software packages CIAO 3.2 (Fruscione et al. 2006), XSPEC 12.2.1
and XRONOS 5.21 (Arnaud 1996; Blackburn 1995). For spectral
analysis, background subtracted spectrum was created using the CIAO
tools and the spectrum was binned such that each bin contained data
with signal-to-noise ratio higher than 3. The data were then fitted
with XSPEC. In general, data bins below 0.3 keV and above 5.0 keV were 
omitted
due to low statistical quality. For the timing analysis, data times
were barycentrically corrected and a background subtracted light
curve was extracted using CIAO.  The light curve was then folded.
The phase resolved spectroscopy was performed by extracting spectra
using the appropriate phases with the $\cha$ tools, then fitted
using XSPEC. We also used CIAO 4.0 Beta 2 version and performed $ACIS 
process events$ and checked if differences in spectra and light curve 
existed. We found no significant changes in the results.

The original observation of RR Pic was conducted in order to search
for the extended emission from the old nova shell in accordance with the
X-ray remnant of the old nova GK Persei (Balman 2005).
Balman (2006) show that an extended emission is marginally detected in a
short exposure with a count rate of 0.0023 $\cs$ above the background as
an elongated region extending from N to S about $\pm$19 arcsec.
The elongation in the $\cha$ image is in agreement with the
orientation of the equatorial
ring in the H$_{\alpha}$ image. Images can be found in Balman (2006) and
 some preliminary analysis of the $\cha$ observation of RR Pic is in
Balman and K\"upc\"u-Yolda\c{s} (2004).

\section{The $\cha$ Spectrum of the Point Source}

The $\cha$  spectrum of the source can not be fitted by a single or
two-temperature bremsstrahlung model including neutral Hydrogen absorption
({\it phabs}, {\it wabs} (models in XSPEC))
with a reduced $\chi^2$ value smaller than 2. A fit with a single
bremsstrahlung model is presented in Figure 1a. The fit doesn't actually
reflect the features of
the spectrum properly, leaving a hard excess above 2 keV. Moreover, the
residuals of the fit in Figure 1a are scattered around the mean up to
4$\sigma$  in the 0.5-2 keV
region. To reduce the scattering of the residuals we added two
absorption lines to the bremsstrahlung model around 1.15 and 1.25 which
improved the fit diminishing the
reduced $\chi^2$ down to 1.75 but still maintaining the scattering
around 0.5-1 keV and
the hard excess (see Figure 1b). Thus, we concluded that there are
line contributions to the continuum in the 0.5-1 keV energy range.
Including emission lines to the fit both reduced the hard excess
and the scattering of the data as well as improving the reduced
$\chi^2$ to 1.13.

Therefore a proper
fit is with a composite model of bremsstrahlung together with a model of
photoelectric absorption of HI, two absorption lines around 1.1-1.4 keV
and 5 Gaussians
centered on the likely emission lines around
0.3-1.1 keV energies. The fit is  shown in Figure 1c and
spectral parameters are shown in Table 1.
The Gaussian lines fitted
to the spectrum correspond to a range of emission lines of Fe (transitions
between XVII and XXIV), S (XIV and XVI), Ca (transitions between XV and
XVII), Ne (IX, X), O (VII, VIII), C VI. The absorption lines correspond to
Fe (transitions between XVII and XXIV), Ne (X, IX) and Na X. However,
due to the spectral resolution limitations of ACIS-S,
the precise energy of the emission lines and absorption lines cannot be
determined.

According to Mukai et al.
(2003) X-ray spectra of CVs can be categorized as being of a
cooling flow plasma emission model or of a
photoionized plasma emission model.
The source spectrum can not be fit 
with the photoionization model
$photoion$
\footnote{\scriptsize See
http://heasarc.gsfc.nasa.gov/docs/xanadu/xspec/models/photoion.html}
 so the shocked plasma emission
can be inferred to be a cooling flow gas as in other nonmagnetic CVs V603 Aql, U Gem and
SS Cyg (Mukai et al. 2003). Therefore, we fitted the spectrum with
a cooling flow emission model
with variable abundances such as VMCFLOW (Mushotzky \& Szymkowiak
1988). The model fits the spectrum well with a reduced $\chi^2$  of
1.58 showing increased abundance ratios of He, C, N and O with respect 
to solar (see Figure 1f for the fit and Table 1 for the spectral
parameters).

Adding a blackbody component along with the bremsstrahlung or other
plasma models as suggested by
van Teeseling et. al. (1996) improves the fits;  however the blackbody 
temperature derived from the
fits are around 0.4-0.7 keV which is too high for any physical 
interpretation.
The blackbody temperature should be in a range 0.01-0.06
keV for mCVs (e.g. Evans $\&$ Hellier 2007). Also, such a range of   
blackbody
temperatures are not typical of nonmagnetic CVs (Baskill et al. 2005).

\section{Temporal Analysis of the Point Source }

The background subtracted light curve extraction of the source was
done by the standard CIAO ACIS light curve extraction procedures.
Plotting and epoch folding were performed with XRONOS 5.21.

Figure 2a shows background subtracted time series with bin time of
1100 s. The X-ray light curve of the source shows variation when
folded on the orbital period of the system, 12528 s (0.14502545(7) d)
which is shown in Figure 2b. The light curve in Figure 2a has been
used for the folding process. The ephemeris of the radial velocities
from Schmidtobreick et al.
(2003) was used (HJD$_{start}$ = 2452328.578335+0.14502545$E$) in the folding process. A simple
sinusoidal fit to the folded light curve yields an X-ray modulation
amplitude of 0.02 $\cs$ with 33\% statistical error.

We performed orbital phase resolved spectroscopy of the source
by extracting phases
between 0-0.2 for the maxima and 0.3-0.45 for the minima. Figure 3a
shows the fit to the maximum spectrum and 3b the minimum spectrum
using a model with a single bremsstrahlung together with a model for
photoelectric absorption of HI. The gaussian absorption and emission lines
used in the entire spectrum were not used in these spectra.
The low statistical quality of the minimum and maximum phase
spectra does not allow the detection of lines. The spectral parameters
are given in Table 2. The
fits show a difference in the column density of neutral Hydrogen
absorption between minimum and maximum
spectrum in the 3$\sigma$ confidence range.

\begin{table}
\centering \label{1} \caption{Spectral parameters of maximum and
minimum spectra in the 0.3-5 keV region. Both spectra were fitted
with a model of bremsstrahlung and photoelectric absorption of HI.
$N_{H}$ is the absorbing column, $kT_{Bremss}$ is the bremsstrahlung
temperature and $K_{Bremss}$ is the bremsstrahlung normalization.
Error ranges for $kT_{Bremss}$ and $K_{Bremss}$ correspond to
2$\sigma$\ confidence level, error range for $N_{H}$ correspond to
3$\sigma$\ confidence level.}

\begin{tabular}{@{}llrrrrlrlr@{}}

\hline
\hline

\multicolumn{1}{l}{}  &
\multicolumn{1}{l}{Maxima} &
\multicolumn{1}{l}{Minima}\\

\hline

N$_{H}$ ($\times$ 10$^{22}$ atoms/cm$^{2}$) &  0.25$^{+0.23}_{-0.18}$ &
0.64$^{+0.14}_{-0.13}$ \\
kT$_{Bremss}$ (keV) & 0.35$^{+0.2}_{-0.13}$ & 0.14$^{+0.21}_{-0.04}$ \\
K$_{Bremss}$ & 0.0004$^{+0.0028}_{>}$ &
0.03$^{<}_{-0.03}$ \\
$\chi^2_{\nu}$ & 1.05 (15 d.o.f.) & 0.49 (6 d.o.f.) \\

\hline
\end{tabular}
\end{table}

\begin{figure*}
\centerline{
\includegraphics[scale=0.35,angle=270]{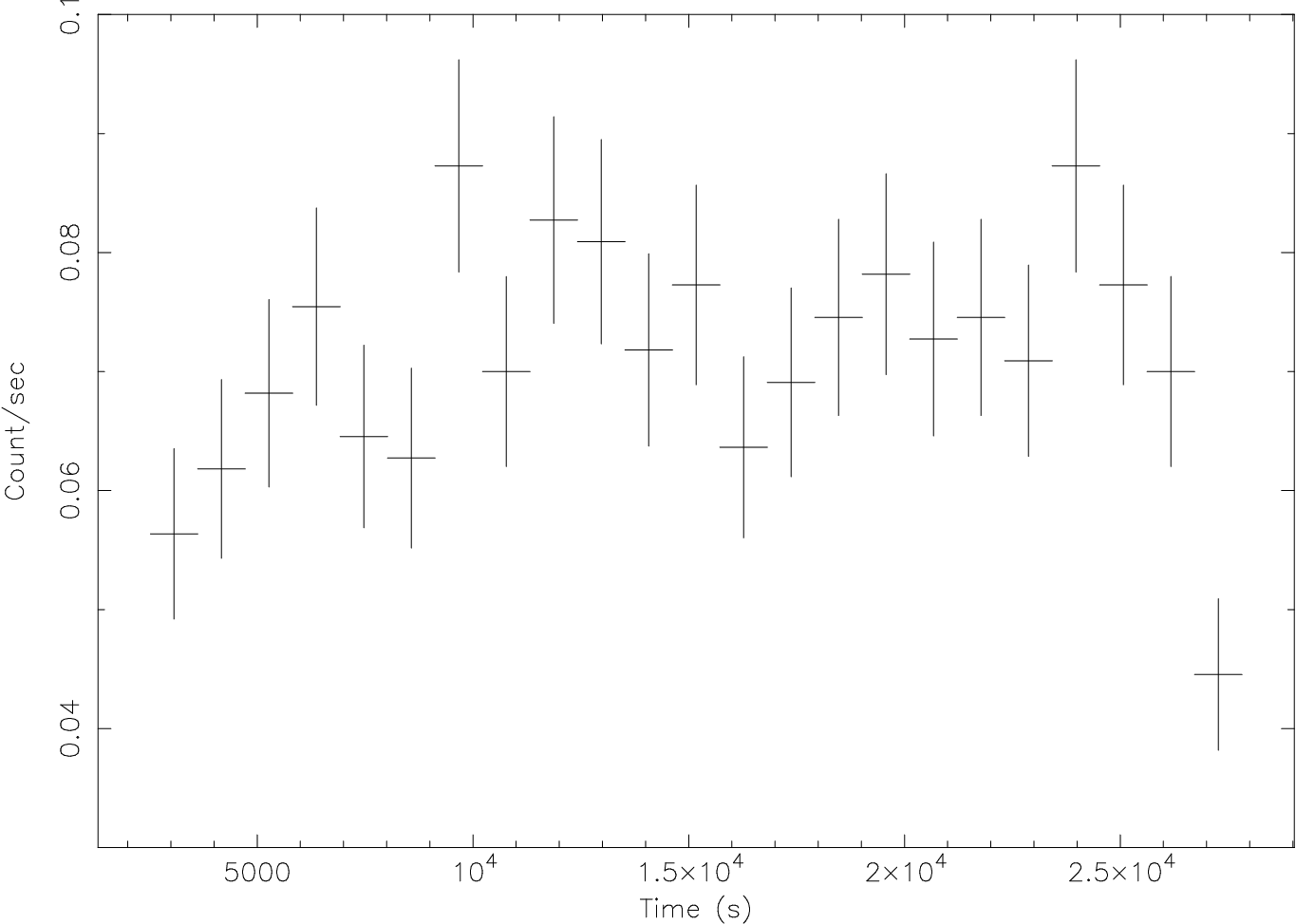} 
\includegraphics[scale=0.35,angle=270]{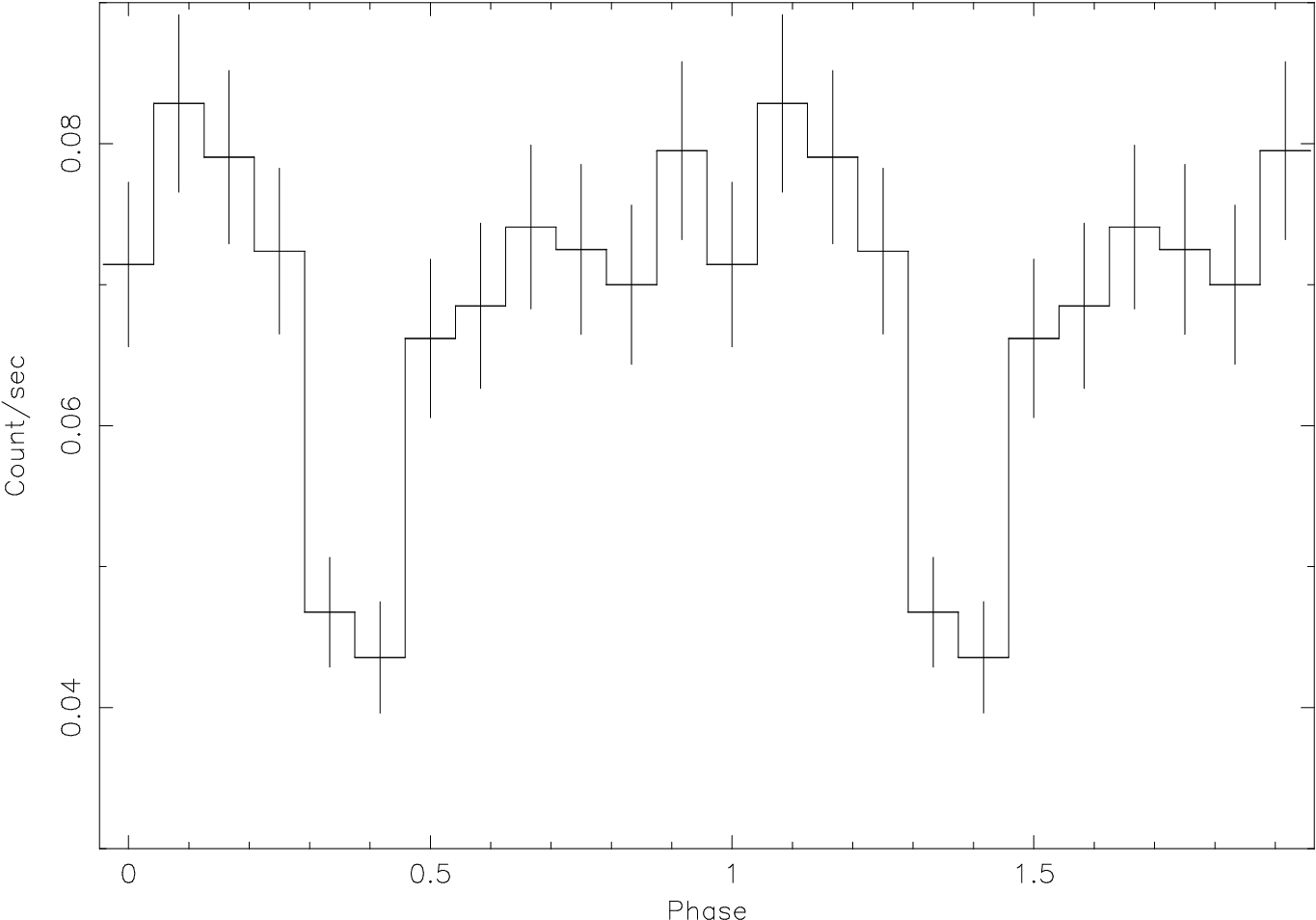}}
\caption{Left-hand panel is the light curve of RR Pic with a bin time of 
1100 s.  Right-hand panel is the light
curve of RR Pic folded over the orbital period of 0.145025 days.}
\end{figure*}

\begin{figure*}
\centerline{
\includegraphics[scale=0.35,angle=270]{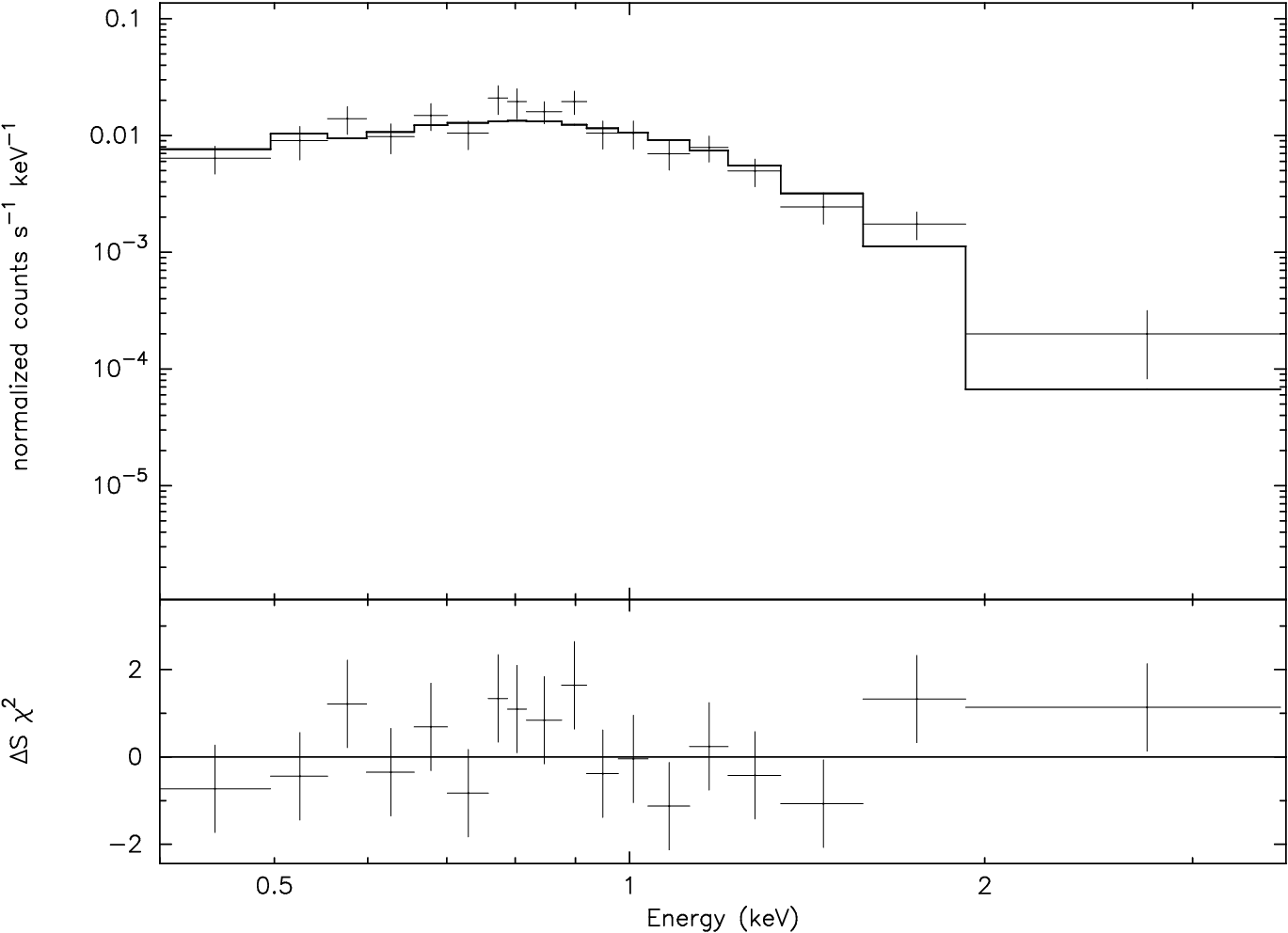} 
\includegraphics[scale=0.35,angle=270]{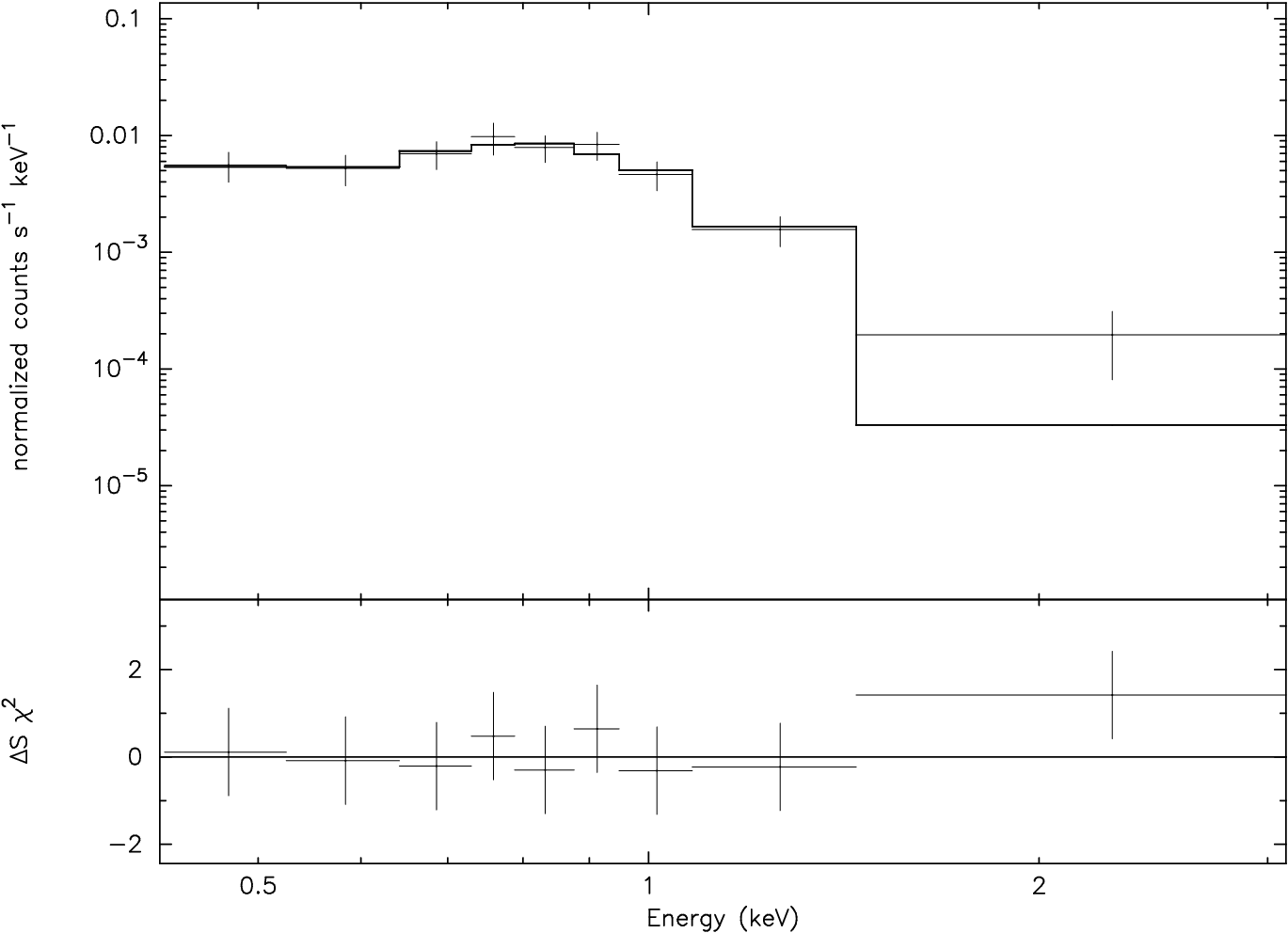}} 
\caption{The spectra of orbital
phase maximum on the right-hand panel and minimum on the left.
Both spectra were  fitted with a model of bremsstrahlung and photoelectric
absorption of HI. The crosses show the data with error
bars and solid lines show the composite
fitted model.
The panels under the spectra show residuals in standard deviations.}
\end{figure*}

\section{Discussion}

 The $\cha$ ACIS-S spectrum of RR Pic fits well with two spectral
models. First one is a composite
model of single bremsstrahlung, photoelectric absorption, two
absorption lines and 5 Gaussian lines at 2-3$\sigma$ confidence
level. Another good fit is with VMCFLOW model with enhanced
abundances of
He, C, N, O and Ne. The good fitting models are consistent with each
other since the Gaussian lines of the composite model are also in
accordance with the abundances of VMCFLOW.

The X-ray luminosity can be calculated as (1.1$\pm$0.2)$\times$10$^{31}$
erg s$^{-1}$ (using the relation L=4$\pi$d$^{2}$F). Assuming the accretion
luminosity is half the total luminosity;
taking the mass and the radius of the white dwarf to be 1 M$_{\odot}$ and
 10$^{9}$ cm, respectively; the accretion rate  $\dot{M}$ is 
1.7$\times$10$^{14}$ g s$^{-1}$ (using the relation L=G$\dot{M}$M/2R). 
These values are
consistent with those of the nonmagnetic CVs in quiescent state
(Baskill et al. 2005).

The fit residuals, in general, show some excess (particularly with VMCFLOW model)
on the harder energy part of the spectrum. In 
order to understand the nature of the excess, spatially resolved spectra 
around the position  of the source 
was extracted from the south-southeast and north-northwest regions 
of the source in accordance with the results of Balman 2002 and 2006. 
Both regions fit well with the composite model noted in Table 1 of 
bremsstrahlung and emission lines. However, a 2-3 $\sigma$ 
hard excess in the south-southeast region needs an additional bremsstrahlung 
component to be added to decrease the reduced $\chi^2$ values down to desirable levels
below 2.  This component has very high absorption around 
50$\times$10$^{22}$ cm$^{-2}$
and plasma temperature in a range 0.36-0.70 keV. This excess emission
can be attributed to emission from the shocked nova shell. In order to 
investigate this using the Rankine-Hugoniot
jump conditions one can derive a relation between the post-shock temperature 
and expansion speed of the material (nova ejecta or the ring) as
kT$_{shock}$=(3/16)$\mu$m$_H$(v$_{shell}$)$^2$. Taking v$_{shell}$ as 
400-850 km s$^{-1}$ range for RR Pic (see Introduction), the shocked nova shell 
temperature is 0.31-1.4  keV consistent with our derived fit result
suggesting a component from the nova shell. Moreover, the high absorption
towards this component can be expected if it is of the reverse shock, 
high absorption in the reserve shock was also detected for the shell of GK Persei 
(Balman 2005).

The abundances of He, C, N and O derived from the fits to the source
spectrum with the VMCFLOW model
are significantly enhanced, where fits with solar abundances are
ruled out with reduced $\chi^2$ larger than 2. This means that the
X-ray emitting region in the point source
consists of enhanced abundances of these elements which can be
explained by the mixing that has occurred in this system.
Prior to the thermonuclear runaway in a classical nova explosion,
diffusion by convection takes place between the hydrogen rich
boundary layer and the white dwarf where heavier elements such as
He, C, N and O are present (Livio 1994). Hence, as heavier elements
diffuse into the boundary layer, the abundances are increased. What
we observe in the spectrum of RR Pic can be a very rare example of
diffusive mixing in a Classical Nova.

The light curve of the source shows variation folded on the orbital
period with a modulation amplitude of 0.02 counts s$^{-1}$ with $\%$33
statistical error. Comparing the spectra of the orbital maximum and
minimum of the source reveals that the X-ray modulation could be due
to the change in the difference in photoelectric absorption between
the two phases. In order to explain this, an analogy to the Low Mass
X-ray Binaries (LMXBs) can be used. In some LMXBs, the X-ray
intensity shows dipping behaviour during the  orbital motion, caused
largely by the obscuration of the X-rays by the region where the
accreting material impacts the disc (White $\&$ Swank 1982). Thus in
our case, the X-ray modulation can be due to the obscuration by a
warm/cold region on the disc, similar to LMXBs. As Parker, Norton
and Mukai (2005) states, a similar orbital modulation is also
present in some CVs. Their work suggests that decrease of modulation
depth with increasing energy implies photoelectric absorption due to
material at the edge of the disc for the systems they have studied.
Furthermore, according to the doppler tomography analysis of RR Pic
by Schmidtobreick et al. (2003), a strong emission region is
detected at phase 0.3 opposite  the bright hot spot (i.e., accretion
impact region). Using their radial velocity ephemeris, we find that
the X-ray eclipse/dip corresponds to the same orbital phase 0.3.
Therefore this supports the idea that the X-ray modulation is due to
an absorbing region on the disc.

The two absorption features that are detected in the spectrum
of RR Pic around 1.1-1.2 keV
are not accounted for before in any CV. We also checked the data 
against any instrumental feature or
processing differences between software versions and the features are
persistent.They can be
fitted with absorption lines
corresponding to Ne IX and Fe (transitions between XVII--XXII) for the
first, and Ne X and Fe (transitions between XIV--XXVI) for the second
feature. These features could be caused by a region
on the disc or the shell itself in the line of sight. Further investigation of
the spectrum could prove useful in finding the origin of these
features.

\section{Summary and Conclusions}

We present the first broad X-ray spectrum of RR Pic in the 0.2-10.0 keV
range using Chandra ACIS-S3 observations.

The spectrum of the source shows a bremsstrahlung temperature in the
range 1-1.6 keV that is consistent with the
temperature range of the non-magnetic CVs of 1-5 keV (Kuulkers
et. al 2006). Emission lines of the elements Fe (transitions between
XVII and XXIV, (Fe L complex)), S (XIV and XVI), Ca (transitions
between XV and XVII), Ne (IX, X), O (VII, VIII), C VI; and
absorption lines of elements Fe (transitions between XVII and XXIV),
Ne (X, IX) and Na X are consistent with the spectrum. 
In order to determine the exact transitions of the emission 
and absorption lines, observations with higher sensitivity and resolution is needed.

The absorption features detected in the spectrum are particularly 
important since no 
absorption feature at these energies were detected before in any CV. So a 
secure detection/confirmation of the absorption lines would be a novel 
discovery if further observations are carried out.

The spectrum clearly fits well with VMCFLOW model rather than the $photoion$ 
model which implies that the X-rays from the system arises from a shocked 
cooling flow plasma. From the VMCFLOW fit, we derive temperature range 
with highT=1.8$^{+0.2}_{-0.2}$ keV and lowT=0.14$^{+0.10}_{>}$ keV. The 
fit also reveals high abundance ratios 
with respect to the solar abundances such as  18.4$^{+9.1}_{-6.1}$, 
1.7$^{+16.5}_{>}$, 8.7$^{+7.1}_{>}$ and 1.9$^{+1.1}_{-0.6}$ for 
He, C, N and O respectively. These abundance enhancements indicate 
existing mixing in 
the boundary layer. The high He and N abundances indicate 
Hydrogen burning which would be a consequent result of a nuclear burning 
stage of a classical nova explosion.  So the spectrum of RR Pic demonstrates 
a very rare example of diffusive mixing prior to and after 
the thermonuclear runaway 
in a Classical Nova detected long years after the explosion. We also find
excess emisison in the spectra (2-3$\sigma$) that we attribute the origin
to be the shocked nova shell with a temperature of 0.3-0.7 keV.

We clearly detect the orbital modulation of RR Pic in the X-ray
wavelengths. Using the ephemeris given by Schmidtobreick et al. (2003), 
the phase of the X-ray eclipse of the system overlaps with the phase of the emission 
region from the disc opposite of the bright spot. 
We also find that the neutral Hydrogen column density
differs for orbital
minimum and orbital maximum spectra with values
(0.25$^{+0.23}_{-0.18}$)$\times$10$^{22}$ \nhu and 
(0.64$^{+0.13}_{-0.14}$)$\times$10$^{22}$ \nhu at 3$\sigma$\ confidence 
level.
Therefore we favor a model where a 
possible warm/cold absorbing region on the disc or line of sight, the nova shell, 
modifiying the spectrum.

\section*{Acknowledgments}

We thank the anonymous referee for critical reading of the manuscript.

YP acknowledge support from
T\"UB\.ITAK, The Scientific and Technological Research Council
of Turkey,  through project 106T040. SB acknowledges
an ESA fellowship and also a TUBA-GEBIP fellowship from the
Turkish Academy of Sciences.

\appendix

\bsp

\label{lastpage}

\end{document}